\documentstyle[12pt,titlepage]{article}

\font\grande=cmr9.5 scaled \magstep4
\font\medio=cmr9.5 scaled \magstep2
\outer\def\beginsection#1\par{\medbreak\bigskip
      \message{#1}\leftline{\bf#1}\nobreak\medskip
\vskip-\parskip
      \noindent}
\def\laq{\raise 0.4ex\hbox{$<$}\kern -0.8em\lower 0.62
ex\hbox{$\sim$}}
\def\gaq{\raise 0.4ex\hbox{$>$}\kern -0.7em\lower 0.62
ex\hbox{$\sim$}}

\begin{document}
\bibliographystyle {unsrt}

\titlepage

\begin{flushright}
CERN-PH-TH/2014-015
\end{flushright}

\vspace{15mm}
\begin{center}
{\grande Scaling laws and sum rules for the B-mode polarization}\\
\vspace{15mm}
 Massimo Giovannini 
 \footnote{Electronic address: massimo.giovannini@cern.ch} \\
\vspace{0.5cm}
{{\sl Department of Physics, Theory Division, CERN, 1211 Geneva 23, Switzerland }}\\
\vspace{1cm}
{{\sl INFN, Section of Milan-Bicocca, 20126 Milan, Italy}}
\vspace*{2cm}

\end{center}

\vskip 1.5cm
\centerline{\medio  Abstract}
The formation of the microwave background polarization anisotropies is investigated
when the stochastic Faraday rate is stationary, random and Markovian. 
The scaling properties of the polarization power spectra and of their nonlinear combinations 
are scrutinized as a function of the comoving frequency.
It is argued that each frequency channel of a given experiment measuring 
simultaneously the E-mode and the B-mode spectra can be analyzed 
in this framework with the aim of testing the physical origin of the polarization
 in a model-independent perspective.
\noindent

\vspace{5mm}

\vfill
\newpage
Synchrotron sources are known to emit polarized radiation \cite{SYNC1} that is stochastically rotated by the Faraday effect \cite{SYNC1,SYNC2}.
To obtain a suitable physical description of the frequency scaling, the corresponding polarization observables are customarily averaged over the rotation rate \cite{SYNC2}. Unlike the case of synchrotron emission, the degree of linear polarization of the cosmic microwave background (CMB in what follows) stems directly from the adiabatic initial conditions of the Einstein-Boltzmann hierarchy implying that the position of the first acoustic peak in the TT correlations\footnote{Following the established terminology, the B-mode autocorrelations are denoted by BB. With similar logic we shall mention throughout the text the TT, TE and EE angular power spectra denoting, respectively, the autocorrelations of the temperature, the autocorrelations of the E-mode and their mutual cross-correlations.} must be roughly $4/3$ times larger than the position of the first anticorrelation peak in the TE angular power spectrum. This prediction has been observationally established by the various data releases of the WMAP collaboration \cite{WMAP9} and later confirmed by independent polarization experiments \cite{ex}.  
The aim of this investigation is to discuss the formation of the CMB polarization by characterizing the Faraday rate as a stationary and random process with approximate Markovian behaviour. Within this novel approach, exploiting the analogies with the stochastic rotation of the synchrotron polarization,  various scaling properties of the corresponding angular power spectra can be derived and eventually tested if and when multifrequency measurements of the B-mode will be available.

We shall consistently work in a conformally flat space-time whose metric tensor can be written as 
$g_{\mu\nu} = a^2(\tau) \eta_{\mu\nu}$  where $\eta_{\mu\nu}$ is the Minkowski metric; $a(\tau)$ shall denote the scale factor and $\tau$ is the conformal time coordinate.
With these necessary specifications, in the concordance paradigm the Faraday rotation rate can be expressed as:
\begin{equation}
X_{F}(\vec{x},\tau)= \frac{\overline{\omega}_{Be}}{2} \biggl(\frac{\overline{\omega}_{pe}}{\overline{\omega}}\biggr)^2 =  \frac{e^3}{ 2 \pi} \biggl( 
\frac{n_{e}}{m_{e}^2 \, a^2 }\biggr)\, \biggl( \frac{\vec{B} \cdot \hat{n}}{\overline{\nu}^2}\biggr), 
\label{rot1}
\end{equation} 
where $\overline{\omega}= 2 \pi \overline{\nu}$ is the (comoving) angular frequency while $\overline{\omega}_{Be}$  and $\overline{\omega}_{pe}$ denote the comoving Larmor and plasma frequencies. The WMAP experiment observes the microwave sky in five frequency channels ranging from $23$ GHz to $94$ GHz.
The Planck satellite explores instead the microwave sky in nine frequency channels: three of them are at low frequency (between $30$ and $70$ GHz) while the remaining six are located between $100$ and $857$ GHz. As it can be swiftly verified from Eq. (\ref{rot1}), the actual value of $X_{F}$ is not necessarily much smaller than $1$ but it can be ${\mathcal O}(1)$ for comoving field strengths of few nG (i.e. $1\, \mathrm{nG} = 10^{-9}\, \mathrm{G}$) and frequencies ${\mathcal O}(10)$ GHz.

The Faraday rate introduced in Eq. (\ref{rot1}) enters directly the evolution of the magnetized brightness perturbations (see e.g.\cite{far1,rev} and references therein) 
whose explicit form in the conformally flat case is given by:
\begin{equation}
\Delta_{\pm}' + (\epsilon' + n^{i} \, \partial_{i} ) \Delta_{\pm} = {\mathcal M}(\vec{x},\tau) \mp 2 i \, X_{F}(\vec{x},\tau) \Delta_{\pm},
\label{re3}
\end{equation}
where $\Delta_{\pm}(\vec{x},\tau)  = \Delta_{Q}(\vec{x},\tau) \pm 
i \, \Delta_{U}(\vec{x},\tau)$; $\Delta_{Q}(\vec{x},\tau)$ and $\Delta_{U}(\vec{x},\tau)$ define the brightness perturbations of the corresponding Stokes 
parameters. In Eq. (\ref{re3}) the prime denotes a derivation with respect to the conformal time coordinate $\tau$; $\epsilon' = \tilde{n}_{e} \, x_{e} \, \sigma_{e\gamma}\, a(\tau)$ is the differential optical depth and the source term ${\mathcal M}(\vec{x},\tau)$ is determined by the electron-photon scattering cross section $\sigma_{e\gamma}$ and by the properties of the magnetic field. The comoving and the physical electron concentrations 
appearing, respectively, in  Eq. (\ref{rot1}) and in the definition of the differential optical depth $\epsilon'$ are related as $n_{e} = a^3(\tau)\, \tilde{n}_{e}$.
The conventional discussion assumes that first the polarization is formed and then it is Faraday rotated with 
 $X_{F} \ll 1$, as it happens if the ambient magnetic field is not too high and the observational frequency is not too small. The goal of the latter approach is to derive a set of phenomenological bounds on the comoving magnetic field $\hat{n}\cdot\vec{B}$ that must be, a priori,  smaller than the nG to comply with the assumed smallness of the Faraday rate. 
 
Rather than deriving a further bound of the magnetic field intensity the purpose here is to explore a different approach where the Faraday rate is described as a random, stationary and approximately Markovian process. The randomness implies that $X_{F}(\tau)$ is not a deterministic variable but rather a stochastic process which is stationary insofar as the autocorrelation function $\Gamma(\tau_{1},\tau_{2}) = \langle X_{F}(\tau_{1}) X_{F}(\tau_{2}) \rangle$ only depends on time differences i.e. $\Gamma(\tau_{1},\tau_{2}) = \Gamma(|\tau_{1}- \tau_{2}|)$; furthermore we shall also assume that the process has zero mean, even if this 
is not strictly necessary for the consistency of the whole approach. 
If  $\tau_{b}$ defines the time-scale of variation of the brightness perturbations of the polarization observables, the physical 
situation investigated here corresponds to $\tau_{b} \gg \tau_{c}$ where $\tau_{c}$ is the correlation time-scale of $X_{F}$.
In the simplest case of a Gaussian-correlated process the autocorrelation function $\Gamma(\tau_{1} - \tau_{2}) = F(\tau_1) \tau_{c} \delta(\tau_{1} - \tau_{2})$.  
If the time scale of spatial variation of the rate is comparable with the time scale of spatial variation of the gravitational fluctuations, 
$X_{F}$ can be considered only time dependent (i.e. a stochastic process). In the opposite situation the Faraday rate must be considered fully inhomogeneous (i.e. a stochastic field). These two possibilities will be separately considered hereunder. On a purely logical ground  
$X_{F}$  can just be a random variable characterized by a given probability distribution and this 
is somehow the most naive case that has been already analyzed in the framework of the synchrotron emission (see, e.g. the second paper of Ref. \cite{SYNC2}) and that will not be treated here.

If $X_{F}(\tau)$ is interpreted as a stochastic process, Eq. (\ref{re3})  becomes, in Fourier space,
\begin{equation}
\delta_{\pm}' + (i k \mu + \epsilon') \delta_{\pm} =\frac{3}{4}(1 -\mu^2) \epsilon' S_{P}(\vec{k}, \tau) \mp 2 i  \, X_{F}(\tau) \delta_{\pm},
\label{un2}
\end{equation}
where $S_{P}(\vec{k},\tau) = (\delta_{I\,2} + \delta_{P\,0} + \delta_{P\,2})$ and $\delta_{\pm}(\vec{k}, \tau)$ denotes the Fourier transform of $\Delta_{\pm}(\hat{n},\tau)$;  
$\delta_{P0}$ and $\delta_{P2}$ are the monopole and the quadrupole of $\delta_{P}$ and  $\delta_{I2}$ is the 
quadrupole of the brightness perturbation related to the intensity of the radiation field.
Equation (\ref{un2}) must be complemented by the evolution of  the brightness perturbations of the intensity (i.e. $\delta_{I}$) that can be used to solve approximately the system in the tight-coupling limit \cite{nase}. 
The source term $S_{P}(\vec{k},\tau)$ depends on the frequency of the channel since the magnetic field 
modifies the trajectories of the electrons scattering the CMB the photons; for sake of simplicity this effect (that is also frequency dependent) shall be neglected in what follows but it is described in detail in the last paper of Ref. \cite{far1} and it can be easily included. 

For equal times (but for different Fourier modes) the fluctuations of the brightness perturbations are random with power spectrum determined by the  
(nearly scale-invariant) spectrum of (Gaussian) curvature perturbations \cite{WMAP9}. Thus, in the absence of Faraday mixing, 
$\delta_{\pm}$ obeys then a deterministic evolution in time while the spatial fluctuations of the polarization are randomly distributed and fixed by the correlation properties of the adiabatic curvature perturbations. Conversely since $X_{F}(\tau)$ is now treated as a stochastic process, Eq. (\ref{un2}) becomes a stochastic differential equation \cite{stoch1} in time and its formal solution is obtainable by iteration: 
\begin{equation}
\delta_{\pm}(\vec{k},\tau) = \sum_{n=0}^{\infty} \delta_{\pm}^{(n)}(\vec{k},\tau), \qquad \delta_{\pm}^{(0)}(\vec{k},\tau)= \delta_{P}(\vec{k},\tau).
\label{un4}
\end{equation}
Equations (\ref{un2}) and (\ref{un4}) imply the following recurrence relations: 
\begin{eqnarray}
\delta_{P}(\vec{k},\tau) &=& \frac{3}{4}(1 - \mu^2) \int_{0}^{\tau} \, e^{- i k\mu (\tau -\tau_{1})} \,  {\mathcal K}(\tau_{1})\, S_{P}(\vec{k}, \tau_{1}),
\label{un5a}\\
\delta_{\pm}^{(n+1)}(\vec{k},\tau) &=&  \pm 2\, i \, \int_{0}^{\tau} \, 
e^{- i k \mu (\tau -\tau_{1})}\, {\mathcal K}(\tau_{1}) \, X_{F}(\tau_1) \, \delta_{\pm}^{(n)}(\vec{k},\tau_{1}).
\label{un5b}
\end{eqnarray}
The differential optical depth enters directly the visibility function giving the probability that a photon is emitted between $\tau$ and $\tau + d\tau$:
\begin{equation}
{\mathcal K}(\tau_{1})=  \epsilon'(\tau_{1})\, e^{ - \epsilon(\tau_{1}, \tau)}, \qquad \epsilon(\tau_{1},\tau) = 
\int_{\tau_{1}}^{\tau} x_{e} \,\tilde{n}_{e}\, \sigma_{e\, \gamma}\, \frac{a(\tau')}{a_{0}}.
\label{un6}
\end{equation}
The full solution of Eq. (\ref{un2}) is formally expressible as: 
\begin{eqnarray}
\delta_{\pm}(\vec{k},\tau) &=& \frac{3}{4} (1 - \mu^2) \int_{0}^{\tau}  \,e^{- i k\mu (\tau -\tau_{1})}\,  {\mathcal K}(\tau_{1})\, S_{P}(\vec{k},\tau_{1})  \, 
{\mathcal A}_{\pm}(\tau, \tau_1) \,d\tau_{1},
\nonumber\\
{\mathcal A}_{\pm}(\tau, \tau_1) &=& e^{\mp 2 \, i\, \int_{\tau_{1}}^{\tau} X_{F}(\tau') \, d\tau'}.
\label{un7}
\end{eqnarray}
The visibility function adopted for the analytic 
estimates has the approximate shape of a double Gaussian whose first peak arises around last scattering (i.e. for $\tau \simeq \tau_{r}$) while the second (smaller) peak occurs for the reionization epoch at a typical redshift of about $11$ \cite{WMAP9,nase}. The 
finite thickness of the last scattering surface does not affect the ratios between the different combinations of polarization power spectra discussed here so that 
the limit of sudden recombination can be safely be adopted; in this limit the first and more pronounced Gaussian profile tends to a Dirac delta function.

The statistical properties of  ${\mathcal A}_{\pm}$ follow directly from the correlation properties of $X_{F}(\tau)$. If, for instance,  
$X_{F}(\tau)$ obeys a stationary and Gaussian process, for any set of $n$ Faraday rates (characterized by different conformal times) the correlator  $\biggl\langle X_{F} (\tau_{1})\, X_{F}(\tau_{2})\,  .\,.\,.\, X_{F}(\tau_{n}) \biggr\rangle $ vanishes if $n$ is odd; if $n$ is even the same correlator equals:
\begin{equation}
\sum_{\mathrm{pairings}}  \biggl\langle X_{F} (\tau_{1})\, X_{F}(\tau_{2})\biggl\rangle\,\biggl\langle X_{F} (\tau_{3})\, X_{F}(\tau_{4})\biggl\rangle
  .\,.\,.\,\biggl\langle X_{F}(\tau_{n-1})  X_{F}(\tau_{n})\biggr\rangle,
  \label{gaus}
\end{equation}
where the sum is performed over all the $(n-1)!$ pairings. In the Gaussian case, the evaluation of the averages can be performed by first 
doing the standard moment expansion and by the resumming the obtained result. As an example, from the explicit expression of ${\mathcal A}_{\pm}$ it follows, 
that 
\begin{equation}
\langle {\mathcal A}_{\pm}(\tau, \tau_{r})\, {\mathcal A}_{\pm}(\tau, \tau_{r})\rangle = \biggl\langle e^{ \pm 4 i \int_{\tau_{r}}^{\tau} X_{F}(\tau')\, d\tau^{\prime}} \biggr\rangle = \sum_{n=0}^{\infty} \frac{(- 2 \omega_{F})^{n}}{n\,!},
\label{un7a}
\end{equation}
where $\omega_{F}$ is given by:
\begin{equation}
 \omega_{F} = 4 \int_{\tau_{r}}^{\tau} \, d\tau_{1} \,  \int_{\tau_{r}}^{\tau} \, d\tau_{2} \langle X_{F}(\tau_{1}) \, X_{F}(\tau_{2}) \rangle.
 \label{cor1}
 \end{equation}
It follows from Eq. (\ref{cor1}) that even if $X_{F} \leq 1$, $\omega_{F}$ is not bound to be smaller than $1$.

If the stationary process is only approximately Markovian, the result  of Eq. (\ref{un7a}) still holds but in an approximate sense. 
While the standard moment expansion can be formally adopted in specific cases (like the Gaussian one) it cannot be used to provide successive approximations. The reason is that any finite number of terms constitutes a bad representation of the function defined by the whole series. This difficulty is overcome with the use of the  cumulants that are certain combinations of the moments. Dropping the functions and keeping 
only their corresponding arguments we have that the relations between the ordinary moments and the cumulants (denoted by $\langle\langle ... \rangle\rangle$) is $\langle 1\rangle = \langle\langle 1\rangle\rangle$, $\langle 1 \, 2\rangle = \langle\langle 1\rangle\rangle\langle\langle2\rangle\rangle + \langle\langle 1\, 2\rangle\rangle$, $\langle 1\,2\,3 \rangle =  \langle\langle 1\rangle\rangle  \langle\langle 2\rangle\rangle\langle\langle 3 \rangle\rangle +  \langle\langle 1\, 2\rangle\rangle \langle\langle 3 \rangle\rangle  + 
 \langle\langle 3\, 1\rangle\rangle \langle\langle 2 \rangle\rangle + \langle\langle 2\, 3\rangle\rangle \langle\langle 1 \rangle\rangle+ 
 \langle\langle 1\,2\,3\rangle\rangle$ and so on and so forth for the other moments of the cluster expansion. Substituting the naive moment expansion with the cumulant expansion we have that the average of Eq. (\ref{un7a}) is given by
\begin{equation}
\langle {\mathcal A}_{\pm}(\tau, \tau_{r})\, {\mathcal A}_{\pm}(\tau, \tau_{r})\rangle = \exp{\biggl[ \sum_{m=1}^{\infty}   \frac{(\pm 4 i)^{m}}{m!} \int_{\tau_{r}}^{\tau} d\tau_{m}\, \, \biggl\langle\biggl\langle X_{F} (\tau_{1})\, X_{F}(\tau_{2})\,  .\,.\,.\, X_{F}(\tau_{m}) \biggr\rangle \biggr\rangle\biggr]}.
\label{cor4}
\end{equation}
 As firstly suggested by Van Kampen (see Ref. \cite{stoch1}, third and fourth paper) 
in the approximately Markovian case the averages of certain stochastic processes will be given by an exponential whose exponent is a series of successive cumulants of $X_{F}$.  All the cumulants beyond the second are zero in the case of an exactly Gaussian process and 
the result reported in Eq. (\ref{un7a}) is recovered. Since each integrand in (\ref{cor4}) virtually vanishes unless $\tau_{1}$, $\tau_{2}$,..., $\tau_{m}$ are close
together, the only contribution to the integral comes from a tube of diameter of order $\tau_{c}$ along the diagonal in the $m$-dimensional integration space. 
More generally, the $m$-th cumulant vanishes as soon as the sequence of times $\tau_{1}$, $\tau_{2}$,...,$\tau_{m}$ contains a gap large compared to $\tau_{c}$.
The is the reason why, in a nutshell, the concept of cumulant is rather practical also in our case. 
 
As an example of stationary process not delta-correlated consider the case where $\Gamma(\tau_{1} - \tau_{2}) = \langle X_{F}(\tau_{1}) \, X_{F}(\tau_{2}) \rangle$  can take only two values $\overline{x}_{F}^2$ and $ - \overline{x}_{F}^2$ and let us suppose that $X_{F}(\tau)$ has switched an even number of times in the interval between $\tau_{1}$ and $\tau_{2}$ so that $\Gamma(\tau_{1} - \tau_{2}) = \overline{x}_{F}^2$ whereas the correlation 
function gives $- \overline{x}_{F}^2$ if there have been an odd number of switches. If $p(n, \Delta\tau)$ is the probability of $n$ switches in the interval $\Delta\tau = \tau_{1} - \tau_{2}$, it follows that 
\begin{equation}
\Gamma(\Delta\tau) = \overline{x}_{F}^2 \sum_{n = 0,\, 2,\, 4\,...}^{\infty} p(n,\Delta\tau) -  \overline{x}_{F}^2  \sum_{n = 1,\, 3,\, 5\,...}^{\infty} p(n,\Delta\tau)
= \overline{x}_{F}^2 \sum_{n =0}^{\infty} (-1)^{n} \, p(n,\Delta\tau).
\label{dich}
\end{equation}
As the switches are random with average rate $r$, $p(n,\Delta\tau) $ is nothing but a Poisson distribution with mean number of switches $\overline{n} = r \, \Delta\tau $, i.e. 
$p_{n} = \overline{n}^{n} e^{- \overline{n}}/n!$. This means that $\Gamma(\Delta\tau) = \overline{x}_{F}^2 \exp{[ - 2 r\, \Delta\tau]}$. This is an example of dichotomic Markov process \cite{stoch1} applied to the case of stochastic Faraday rate. 

If $X_{F}$ is a stochastic field rather than a stochastic process the discussion in mathematically slightly different but physically equivalent as far 
as the frequency scaling is concerned. More specifically, the evolution equations for $\delta_{\pm}$ will now contain a convolution and can be written as:
\begin{equation}
\delta_{\pm}' + ( i k \mu +\epsilon') \delta_{\pm} = \frac{3}{4} ( 1 - \mu^2) \, \epsilon'\,S_{P}(\vec{k},\tau) \mp i \, b_{F}(\overline{\nu},\tau) \int d^{3} p \,\delta_{\pm}(\vec{k} + \vec{p},\tau) \, n^{i} B_{i}(\vec{p},\tau),
\end{equation} 
where we defined, for convenience, $b_{F}(\overline{\nu},\tau) =  2 \, e^3 n_{e}/[(2\pi)^{5/2} m_{e}^2 a^2(\tau) \overline{\nu}^2]$.
The iterative solution of Eqs. (\ref{un4}) and (\ref{un5a})--(\ref{un5b}) becomes, in this case, 
\begin{eqnarray}
&&\partial_{\tau} \delta_{\pm}^{(0)} + ( i k \mu + \epsilon') \delta_{\pm}^{(0)} = \frac{3}{4}(1 -\mu^2) \epsilon' S_{P}(\vec{k},\tau),
\label{nh1}\\
&& \partial_{\tau} \delta_{\pm}^{(1)} + ( i k \mu + \epsilon') \delta_{\pm}^{(1)} = \mp i \, b_{F}(\overline{\nu},\tau) \, \int d^{3} p \, \delta_{P}(\vec{k} + \vec{p}, \tau) \, n^{i} \, B_{i}(\vec{p},\tau),
\label{nh2}\\
&&  \partial_{\tau} \delta_{\pm}^{(2)} + ( i k \mu + \epsilon') \delta_{\pm}^{(2)} = \mp i \, b_{F}(\overline{\nu},\tau) \, \int d^{3} p^{\prime} \, \delta_{P}(\vec{k} + \vec{p}^{\,\prime}, \tau) \, n^{i} \, B_{i}(\vec{p},\tau) \, n^{j}\, B_{j}(\vec{p}^{\,\prime}, \tau).
\label{nh3}
\end{eqnarray}
To compute the averages we must therefore specify the correlation properties of the Faraday rate. Even if the spatial dependence may reside 
in all the terms contributing to the Faraday rate, it is reasonable to presume that the leading effect may come from the magnetic field whose 
correlation function will then be parametrized as:
\begin{equation}
\langle B_{i}(\vec{q},\,\tau_{1}) \, B_{j}(\vec{p},\,\tau_{2}) \rangle = \frac{2 \pi^2}{p^3} P_{ij}(\hat{p}) \, \overline{P}_{B}(p) \, \Gamma(|\tau_{1} - \tau_{2}|) \, \delta^{(3)}(\vec{q} + \vec{p}),
\label{CI1}
\end{equation}
where $\Gamma(|\tau_{1} - \tau_{2}|) = \tau_{c}\, \delta(\tau_{1} - \tau_{2})$ in the delta-correlated case. In the same approximation exploited before and using Eq. (\ref{CI1}),  $\omega_{F}$ becomes now 
\begin{equation}
\omega_{F} = \frac{8 \overline{b}_{F}^2}{3 \overline{\nu}^4} \, \int \frac{d p}{p} \overline{P}_{B}(p) \, \int_{\tau_{r}}^{\tau} d\tau_{1} \int_{\tau_{r}}^{\tau} d\tau_{2} \frac{\Gamma(|\tau_{1} - \tau_{2}|)}{a^2(\tau_{1}) 
a^2(\tau_{2})},
\label{CI2}
\end{equation}
where the constant $\overline{b}_{F}= b_{F}(\overline{\nu},\tau)a^2(\tau) \overline{\nu}^2$ has been introduced in order to draw special attention 
on the frequency scaling that is the most relevant aspect of Eq. (\ref{CI2}), at least in the present approach.

The dependence of the polarization observables upon $\omega_{F}$ can now be determined. Since $\Delta_{\pm}$ transform as fluctuations of spin-weight $\pm 2$, they can be expanded in terms of spin-$\pm2$ spherical harmonics $_{\pm 2}Y_{\ell\,m}(\hat{n})$, with coefficients  $a_{\pm2,\,\ell\, m}$. The E- and B-modes are, up to a sign, the real and the imaginary 
parts of $a_{\pm 2,\ell\,m}$, i.e. $a^{(E)}_{\ell\, m} = -(a_{2,\,\ell m} + a_{-2,\,\ell m})/2$ and $a^{(B)}_{\ell\, m} =  i(a_{2,\,\ell m} - a_{-2,\,\ell m})/2$. The real-space fluctuations constructed from $a^{(E)}_{\ell\,m}$ and $a^{(B)}_{\ell\,m}$ have the 
property of being invariant under rotations on a plane orthogonal to $\hat{n}$.  They are therefore scalars and must
be expanded in terms of (ordinary) spherical harmonics:
\begin{equation}
\Delta_{E}(\hat{n},\tau) = \sum_{\ell\, m} N_{\ell}^{-1} \,  a^{(E)}_{\ell\, m}  \, Y_{\ell\, m}(\hat{n}),\qquad 
\Delta_{B}(\hat{n},\tau) = \sum_{\ell\, m} N_{\ell}^{-1} \,  a^{(B)}_{\ell\, m}  \, Y_{\ell\, m}(\hat{n}),
\label{int4}
\end{equation}
where $N_{\ell} = \sqrt{(\ell - 2)!/(\ell +2)!}$.  
Within these notations, the EE and BB angular power spectra are defined as 
\begin{equation}
C_{\ell}^{(EE)} = \frac{1}{2\ell + 1} \sum_{m = -\ell}^{\ell} 
\langle a^{(E)*}_{\ell m}\,a^{(E)}_{\ell m}\rangle,\qquad 
C_{\ell}^{(BB)} = \frac{1}{2\ell + 1} \sum_{m=-\ell}^{\ell} 
\langle a^{(B)*}_{\ell m}\,a^{(B)}_{\ell m}\rangle,
\label{int5}
\end{equation}
while the cross-correlation power spectrum is be constructed 
from $\langle a^{(\mathrm{E})*}_{\ell m}\,a^{(\mathrm{B})}_{\ell m} + a^{(\mathrm{E})}_{\ell m}\,a^{(\mathrm{B})*}_{\ell m} \rangle$.
The repeated application of 
generalized ladder operators whose action either raises or lowers the spin weight of a given fluctuation \cite{EB} (see also \cite{far1})
leads to a direct connection between $\Delta_{E}$, $\Delta_{B}$ and $\Delta_{\pm}$ 
\begin{equation}
\Delta_{E} (\hat{n},\tau) = - \frac{1}{2} \partial_{\mu}^2\, \biggl[ ( 1 -\mu^2)  (\Delta_{+} + \Delta_{-})\biggr], \qquad 
 \Delta_{B} (\hat{n},\tau) = \frac{i}{2} \partial_{\mu}^2\, \biggl[ ( 1 -\mu^2)  (\Delta_{+} - \Delta_{-})\biggr],
 \label{BBEE}
 \end{equation}
where $\partial_{\mu}^2$ denotes the second derivative with respect to $\mu= \cos{\vartheta}$. 
From Eqs. (\ref{int4}) and (\ref{BBEE}) we can finally determine $a_{\ell m}^{(E)}$ and $a_{\ell m}^{(B)}$ within the set of conventions 
followed here:
\begin{eqnarray}
a_{\ell m}^{(E)} &=& - \frac{N_{\ell}}{2 (2 \pi)^{3/2}} \int d \hat{n} \, Y_{\ell m}^{*}(\hat{n})\int  d^{3} k \,  \partial_{\mu}^2 \biggl\{ (1 - \mu^2) \biggl[\delta_{+}(\vec{k},\tau) + \delta_{-}(\vec{k},\tau)\biggr]\biggr\},
\nonumber\\
a_{\ell m}^{(B)} &=&  \frac{i\, \,N_{\ell}}{2 (2 \pi)^{3/2}} \int d \hat{n} \, Y_{\ell m}^{*}(\hat{n})\int  d^{3} k \, \partial_{\mu}^2 \biggl\{ (1 - \mu^2) \biggl[\delta_{+}(\vec{k},\tau) - \delta_{-}(\vec{k},\tau)\biggr]\biggr\}.
\label{cor5}
\end{eqnarray}
Inserting Eq. (\ref{cor5}) into Eq. (\ref{int5}), the angular power spectra of the E-mode and of the B-mode polarizations can be derived and they are:
\begin{equation}
C_{\ell}^{(EE)}(\omega_{F}) = e^{- \omega_{F} }\, \cosh{\omega_{F}} \,\overline{C}_{\ell}^{(EE)}, \qquad
C_{\ell}^{(BB)}(\omega_{F}) = e^{- \omega_{F} }\, \sinh{\omega_{F}} \,\overline{C}_{\ell}^{(EE)};
\label{cor6}
\end{equation}
the cross-correlation power spectrum is instead vanishing $C_{\ell}^{(EB)} = 0$. In Eq. (\ref{cor6}) $\overline{C}_{\ell}^{(EE)}$ is the E-mode autocorrelation produced
 by the standard adiabatic mode and in the absence of Faraday mixing.
Equation (\ref{cor6}) shows that the B-mode and the E-mode polarizations are both frequency dependent. 
Despite the fact that these formulas hold also when $\omega_{F} \geq 1$, in the limit $\omega_{F} \ll 1$ 
the standard results are recovered and only the B-mode depends on the frequency \cite{far1}. 
From Eq. (\ref{cor6}) the following sum rules for the angular power spectra can be easily established: 
\begin{eqnarray}
&& C_{\ell}^{(EE)}(\omega_{F}) + C_{\ell}^{(BB)}(\omega_{F}) = \overline{C}_{\ell}^{(EE)},
\label{cor7}\\
&& C_{\ell}^{(EE)}(\omega_{F}) - C_{\ell}^{(BB)}(\omega_{F}) = e^{- 2 \omega_{F}} \overline{C}_{\ell}^{(EE)}.
\label{cor8}
\end{eqnarray}
Introducing now the properly normalized angular power spectra ${\mathcal G}_{E\ell}(\omega_{F})$ and ${\mathcal G}_{B\ell}(\omega_{F})$
\begin{equation}
{\mathcal G}_{E\ell}(\omega_{F})= \frac{\ell (\ell +1)}{2 \pi} \, C_{\ell}^{(EE)}(\omega_{F}), \qquad {\mathcal G}_{B\ell}(\omega_{F}) 
= \frac{\ell (\ell +1)}{2 \pi} \, C_{\ell}^{(BB)}(\omega_{F}),
\label{cor11}
\end{equation}
the following ratio of nonlinear combinations has well defined scaling properties with $\omega_{F}$:
\begin{equation}
{\mathcal L}_{0}(\omega_{F}) = \frac{{\mathcal G}^2_{E\ell}(\omega_{F}) - {\mathcal G}^2_{B\ell}(\omega_{F})}{[{\mathcal G}_{E\ell}(\omega_{F}) + {\mathcal G}_{B\ell}(\omega_{F})]^2}\to e^{- 2 \omega_{F}},
\label{corr12}
\end{equation}
Equation (\ref{corr12}) does not assume that the Faraday rate is much smaller than $1$ and it does not even assume a specific form of the Markov process. For an exactly Gaussian process or for a dichotomic Markov 
process (see e.g. Eq. (\ref{dich})) the explicit expressions of $\omega_{F}$ can be rather different but the frequency dependence will be always the same: since 
$\omega_{F}$ is quadratic in the rates it will always scale as $1/\overline{\nu}^{4} \simeq \lambda^{4}$ where $\lambda$ denotes the wavelength of the channel. Since the scale factor is normalized in such a way that $a_{0}=1$, physical and comoving 
frequencies coincide today but not in the past. The combination reported in Eq. (\ref{corr12}) is not unique 
and different expressions can be envisaged depending on the actual features of the measurement. Two further combinations explicitly depending on $\omega_{F}$ are: 
\begin{equation} 
{\mathcal L}_{1}(\omega_{F}) = \frac{{\mathcal G}_{E\ell}(\omega_{F})  - {\mathcal G}_{B\ell}(\omega_{F})}{{\mathcal G}_{E\ell}(\omega_{F}) +{\mathcal G}_{B\ell}(\omega_{F}) } \to e^{ - 2 \omega_{F}},\quad 
{\mathcal L}_{2}(\omega_{F}) = \frac{{\mathcal G}^2_{E\ell}(\omega_{F})  + {\mathcal G}^2_{B\ell}(\omega_{F})}{{\mathcal G}^2_{E\ell}(\omega_{F}) - {\mathcal G}_{B\ell}^2(\omega_{F}) } \to  \cosh{2\omega_{F}}.
\label{cor10}
\end{equation}
Since ${\mathcal L}_{0}$, ${\mathcal L}_{1}$ and ${\mathcal L}_{2}$ contain ratios of the angular power spectra, the finite thickness of the last scattering surface is not expected to affect these conclusions in any significant manner.  

Stochastic Faraday rotation can be tested through multi-frequency observations once the measurements of the B-mode polarization will be available. 
If the E-mode and the B-mode autocorrelations are independently measured 
in each frequency channel of a given experiment, both scale-invariant and scale-dependent combinations of the angular 
power spectra can be constructed frequency by frequency. So, for instance the combination ${\mathcal L}_{0} + {\mathcal L}_{2} \to 2$ which is scale-invariant in the limit $\omega_{F} \ll 1$. Similarly ${\mathcal L}_{2}/({\mathcal L}_{0} + {\mathcal L}_{0}^{-1})$ is scale-invariant in spite of the value of $\omega_{F}$. Equations (\ref{corr12}) and (\ref{cor10}) illustrate then a possible redundant set of physical observables that can be used to discriminate between the frequency dependence induced by the stochastic Faraday effect or by other concurrent forms of frequency scaling caused either by the known or by the yet unknown foregrounds.  

The present investigation described the Faraday effect of the CMB as a random, stationary and quasi-Markovian process. The stochastic treatment of this physical phenomenon has been explored in analogy with the case of synchrotron polarization. The obtained results encompass and complement previous analyses where the formation of Faraday effect has been customarily presented as a purely deterministic process in time. 
Apart from the discussion of the frequency scaling of the polarization observables, further applications of the approach developed here seem both physically plausible and technically feasible.

\end{document}